\documentclass[12pt]{article}
\newcommand {\evis} {$E_{\mathrm{vis}}$}

\newcommand {\omutau} {$\nu_\mu \rightarrow \nu_\tau$}

\newcommand {\omue} {$\nu_\mu \rightarrow \nu_e$}
\newcommand {\omuebar} {$\bar{\nu}_\mu \rightarrow \bar{\nu}_e$}
\newcommand {\promue} {$P(\nu_\mu \rightarrow \nu_e)$}
\newcommand {\promuebar} {$P(\bar{\nu}_\mu \rightarrow \bar{\nu}_e)$}

\def\nue_pizero{$\nu_e n \rightarrow e^- \pi^0 p$}
\def\anue_pizero{$\bar{\nu}_e p \rightarrow e^+ \pi^0 n$}
\def\ypi_gg{$\pi^0 \rightarrow \gamma\gamma$}

\begin {document}
\title
{Neutrino Superbeams and the Magic Baseline}
\author{
A. Asratyan\thanks{Corresponding author. Tel.: + 7-095-237-0079;
fax: + 7-095-127-0837; {\it E-mail address:} asratyan@vitep1.itep.ru.},
G.V. Davidenko, 
A.G. Dolgolenko,
V.S. Kaftanov,\\
M.A. Kubantsev\thanks{Now at Department of Physics and Astronomy,
Northwestern University, Evanston, IL 60208, USA.},
and V. Verebryusov\\
\normalsize {\it Institute of Theoretical and Experimental Physics,}\\
\normalsize {\it B. Cheremushkinskaya St. 25, Moscow 117259, Russia}
}                                          
\date {\today}
\maketitle

\begin{abstract}
     We examine the sensitivity to \omue\ of a conceptual
experiment with a neutrino superbeam incident on a Megaton-scale
water Cherenkov detector over a "magic" baseline $\sim$7300 km.
With realistic beam intensity and exposure, the experiment may 
unambiguously probe $\sin^2 2\theta_{13}$ and the sign of 
$\Delta m^2_{31}$ down to $\sin^2 2\theta_{13} \sim 10^{-3}$. 
\end{abstract}

     Detecting the subdominant oscillation \omue\ on the "atmospheric" 
scale of $L/E$ has emerged as a priority for long-baseline accelerator
experiments. This is because the \omue\ and \omuebar\ probabilities 
are sensitive to yet-unknown parameters of neutrino mixing: the mixing 
angle $\theta_{13}$, the sign of the "atmospheric" mass-squared 
difference $\Delta m^2_{31}$, and the $CP$-violating phase 
$\delta_{CP}$ \cite{review}. However, extracting the values of these 
parameters from measured probabilities will encounter the problem of 
degenerate solutions \cite{degeneracies}. In particular, the asymmetry
between \promue\ and \promuebar\ may arise from either the intrinsic 
$CP$ violation and the matter effect that is correlated with the sign of  
$\Delta m^2_{31}$ \cite{matter}. The degeneracies can be resolved by 
comparing the data taken with a shorter and longer baselines 
\cite{combine}. Selecting the latter as the "magic" baseline 
$L_\mathrm{magic} \simeq 7300$ km will render this strategy particularly 
effective: for $L = L_\mathrm{magic}$, all  $\Delta m^2_{21}$-induced 
effects like $CP$ violation are predicted to vanish up to second order 
of the small parameter $\Delta m^2_{21} / \Delta m^2_{31}$ 
\cite{degeneracies, magic}. Therefore, 
selecting $L = L_\mathrm{magic}$ may allow to uniquely determine 
$\sin^2 2\theta_{13}$ and the sign of $\Delta m^2_{31}$, but not 
$\delta_{CP}$ which should be probed with a shorter baseline.

     In this paper, we discuss a conceptual experiment that involves
a neutrino "superbeam" incident on a water Cherenkov detector over a
magic baseline of $L = 7340$ km\footnote{This 
is chosen to match the distance from Fermilab to
Gran Sasso or from CERN to Homestake.}.
A water Cherenkov target is selected on the merit of good separation
and spectrometry of electromagnetic showers \cite{resolution}, and is
assumed to be a megaton-scale detector like UNO or Hyper-Kamiokande
\cite{detectors}. In tuning the energy of the neutrino
beam, one must take into account that the $E_\nu$-dependence of
oscillation probability for $L = 7340$ km is strongly affected by 
Earth matter: for $\Delta m^2_{31} > 0$, the matter effect \cite{matter} 
shifts the first maximum of \promue\  down to 
$E_\nu / \Delta m^2_{31} \simeq 2.5\times10^3$ GeV/eV$^2$
from the vacuum value of $5.9\times10^3$ GeV/eV$^2$.
Assuming  $\Delta m^2_{31} = 0.003$ eV$^2$, the oscillation maximum
is at $E_\nu \simeq 7.5$ GeV which conveniently matches the peak of 
$\nu_\mu$ flux in the "Medium-Energy" (or PH2me) beam of Fermilab's Main 
Injector, as designed for the NuMI--MINOS program \cite{numi}. Therefore,
this is selected as the model beam in our simulation. We assume
$1.6\times10^{21}$ protons on neutrino target per year, as 
expected upon the planned upgrade of Main Injector's intensity 
\cite{driver}. In the absence of oscillations, the beam will produce
some 58 $\nu_\mu$CC (21 $\bar{\nu}_\mu$CC) events per 1 kton$\times$yr
in the far detector with the $\nu$ ($\bar{\nu}$) setting of the focusing
system.

     At neutrino energies below 1 GeV, as in the proposed 
JHF--Kamioka experiment \cite{jhf2k}, $\nu_e$ appearance can be
efficiently detected in a water Cherenkov apparatus by selecting 1-ring 
$e$-like events of the reaction  $\nu_e N \rightarrow e^-X$  that is 
dominated by quasielastics. (Here and in what follows, $X$ denotes a 
system of hadrons other than the $\pi^0$, in which the momenta of all 
charged particles are below the Cherenkov threshold in water.) At 
substantially higher energies considered in this paper, using the 1$e$ 
signature of \omue\ is complicated by more background from the 
flavor-blind NC reaction $\nu N \rightarrow \nu \pi^0 X$: its cross 
section increases with $E_\nu$, and so does the fraction of $\pi^0$
mesons whose $\gamma\gamma$ decays produce a single $e$-like ring in 
the water Cherenkov detector\footnote{This
happens when the opening angle is too small for 
the two showers to be resolved \cite{pizero}.}.
In \cite{multi}, we have demonstrated that $\nu_e$ appearance can be 
analyzed with less NC background by detecting the reactions
$\nu_e N \rightarrow e^- \pi^+ X$  and
$\nu_e N \rightarrow e^- \pi^0 X$
that involve emission of a charged or neutral 
pion\footnote{Here and below, corresponding
antineutrino reactions are implicitly
included.}.
We proceed to briefly describe the selections of these CC reactions, 
as formulated in \cite{multi}.

     The reaction  $\nu_e N \rightarrow e^- \pi^+ X$  is selected by
requiring two rings in the detector, of which one is $e$-like and the 
other is non-showering and has a large emission angle of 
$\theta_\pi > 50^0$. This is referred to as the "$e\pi$ signature".
The selection $\theta_\pi > 50^0$ is aimed at suppressing the NC
reaction $\nu p \rightarrow \nu \pi^0 p$ in which the momentum of the 
final proton is above the Cherenkov 
threshold\footnote{This reaction may 
also be rejected by identifying 
relativistic protons by ring shape,
as proposed in \cite{protons}.}.
The residual NC background is largely due to the reaction
$\nu N \rightarrow \nu \pi^0 \pi^\pm X$ with two pions in the final 
state. The $\nu_\mu$CC background arises from the reaction
$\nu_\mu N \rightarrow \mu^- \pi^0 X$ in which the muon is emitted 
at a broad angle. The $\nu_\tau$CC background arises from the dominant 
oscillation \omutau\ followed by  $\nu_\tau N \rightarrow \tau^-\pi^+X$
and  $\tau^- \rightarrow e^- \nu \bar{\nu}$.

     The reaction $\nu_e N \rightarrow e^- \pi^0 X$ is selected by
requiring either three $e$-like rings of which two fit to \ypi_gg, or 
two $e$-like rings that would not fit to a $\pi^0$. This is referred
to as the "multi-$e$ signature". The NC background arises from the 
reaction $\nu N \rightarrow \nu \pi^0 \pi^0 N$ in which 
at least one of the two $\pi^0$ mesons has not been reconstructed.
Note that in the latter reaction the two $\pi^0$ mesons are emitted 
with comparable energies, whereas in  $\nu_e N \rightarrow e^- \pi^0 X$  
the $e^-$ tends to be the leading particle. This suggests a selection
based on the absolute value of asymmetry  $A = (E_1-E_2)/(E_1+E_2)$,
where $E_1$ and $E_2$ are the energies of the two showers for the
two-ring signature, and of the reconstructed $\pi^0$ and the "odd"
shower---for the three-ring signature. In this paper, we use the 
selection $|A| > 0.6$. The $\nu_\tau$CC background is largely due 
to electronic decays of $\tau$ leptons produced in association with
a $\pi^0$. The $\nu_\mu$CC background originates from CC events with
a muon below the Cherenkov threshold and two $\pi^0$ mesons in the 
final state, and is negligibly small.

     In the simulation, the matter effect is accounted for in the 
approximation of uniform matter density along the neutrino path
($\langle\rho\rangle = 4.3$ g/cm$^3$ for $L = 7340$ km), which
adequately reproduces the results of exact calculations for the
actual density profile of the Earth \cite{matter}. Relevant 
neutrino-mixing parameters are assigned the values consistent with
the atmospheric and reactor data \cite{atmo, chooz}:             
$\Delta m^2_{31} = \pm 0.003$ eV$^2$, $\sin^2 2\theta_{23} = 1$, and 
$\sin^2 2\theta_{13} = 0.01$ (the latter value is ten times below
the upper limit imposed in \cite{chooz}). The simulation relies on the
neutrino-event generator NEUGEN based on the Soudan-2
Monte Carlo \cite{neugen}, that takes full account of exclusive    
channels like quasielastics and excitation of baryon resonances.

\begin{figure}
\vspace{18 cm}
\includegraphics{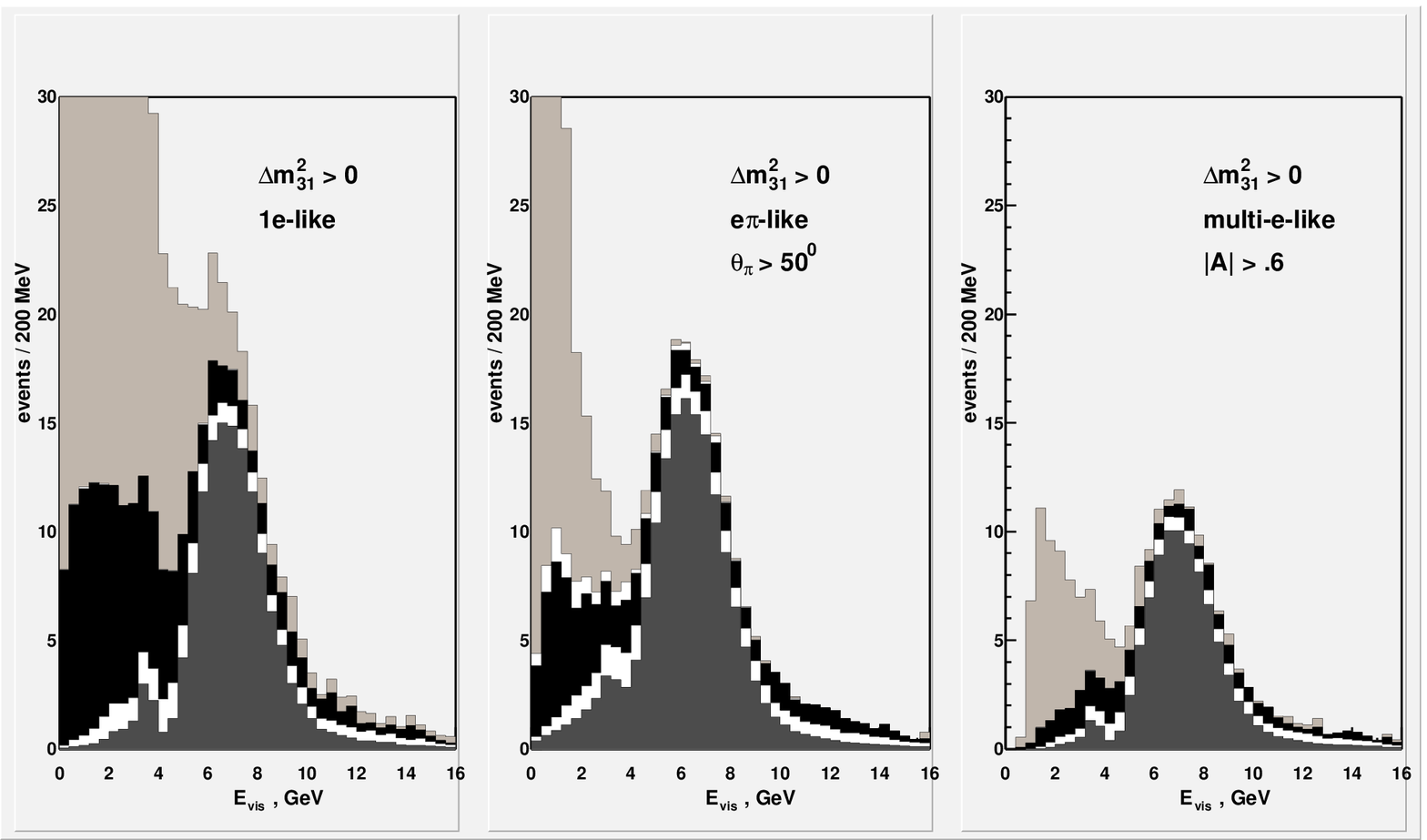}
\caption
{\evis\ distributions of 1$e$-like events (left-hand panel), 
$e\pi$-like events (middle panel), and multi-$e$-like events 
(right-hand panel) for $\Delta m^2_{31} > 0$ and  incident 
neutrinos. From bottom, the depicted components are the \omue\ 
signal (shaded area), intrinsic $\nu_e$CC background (white area), 
$\nu_\tau$CC background (black area), $\nu_\mu$CC background (white
area), and the NC background (light-shaded area).
Event statistics are for an exposure of 1 Mton$\times$yr.}
\label{three}
\end{figure}

\begin{figure}
\vspace{18 cm}
\includegraphics{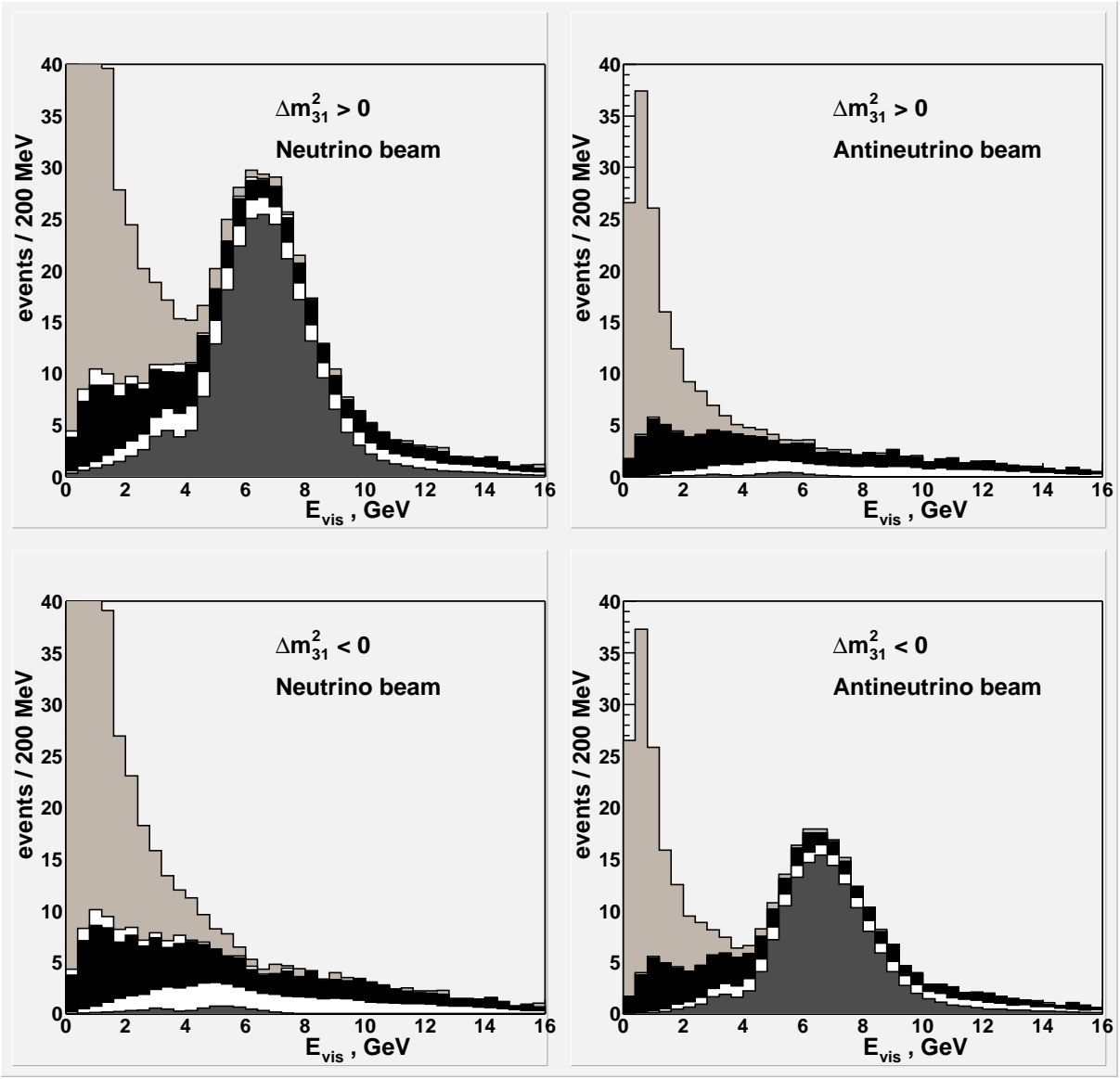}
\caption
{Combined \evis\ distributions of $e\pi$-like and 
multi-$e$-like events for incident neutrinos and antineutrinos
(left- and right-hand panels) and for positive and negative values
of $\Delta m^2_{31}$ (top and bottom panels).
From bottom, the depicted components are the \omue\ 
signal (shaded area), intrinsic $\nu_e$CC background (white area), 
$\nu_\tau$CC background (black area), $\nu_\mu$CC background (white
area), and the NC background (light-shaded area). Event statistics are 
for equal $\nu$ and $\bar{\nu}$ exposures of 1 Mton$\times$yr.}
\label{together}
\end{figure}

    The \evis\ distributions of 1$e$-like, $e\pi$-like, and 
multi-$e$-like events are illustrated in Fig. \ref{three}, assuming 
$\Delta m^2_{31} > 0$ and incident neutrinos. Here, \evis\ stands
for the net energy of all $e$-like rings. Total background to the
\omue\ signal is seen to be the greatest for 1$e$-like events, and
therefore we drop these from further analysis. Combined \evis\ 
distributions of $e\pi$-like and multi-$e$-like events are shown in
Fig. \ref{together} for either beam setting and either sign of
$\Delta m^2_{31}$. With equal $\nu$ and $\bar{\nu}$ exposures of 
1 Mton$\times$yr, the oscillation signal reaches some 250 events for 
$\Delta m^2_{31} > 0$ and incident neutrinos, and some 140 events for 
$\Delta m^2_{31} < 0$ and incident antineutrinos.

\begin{figure}
\vspace{18 cm}
\includegraphics{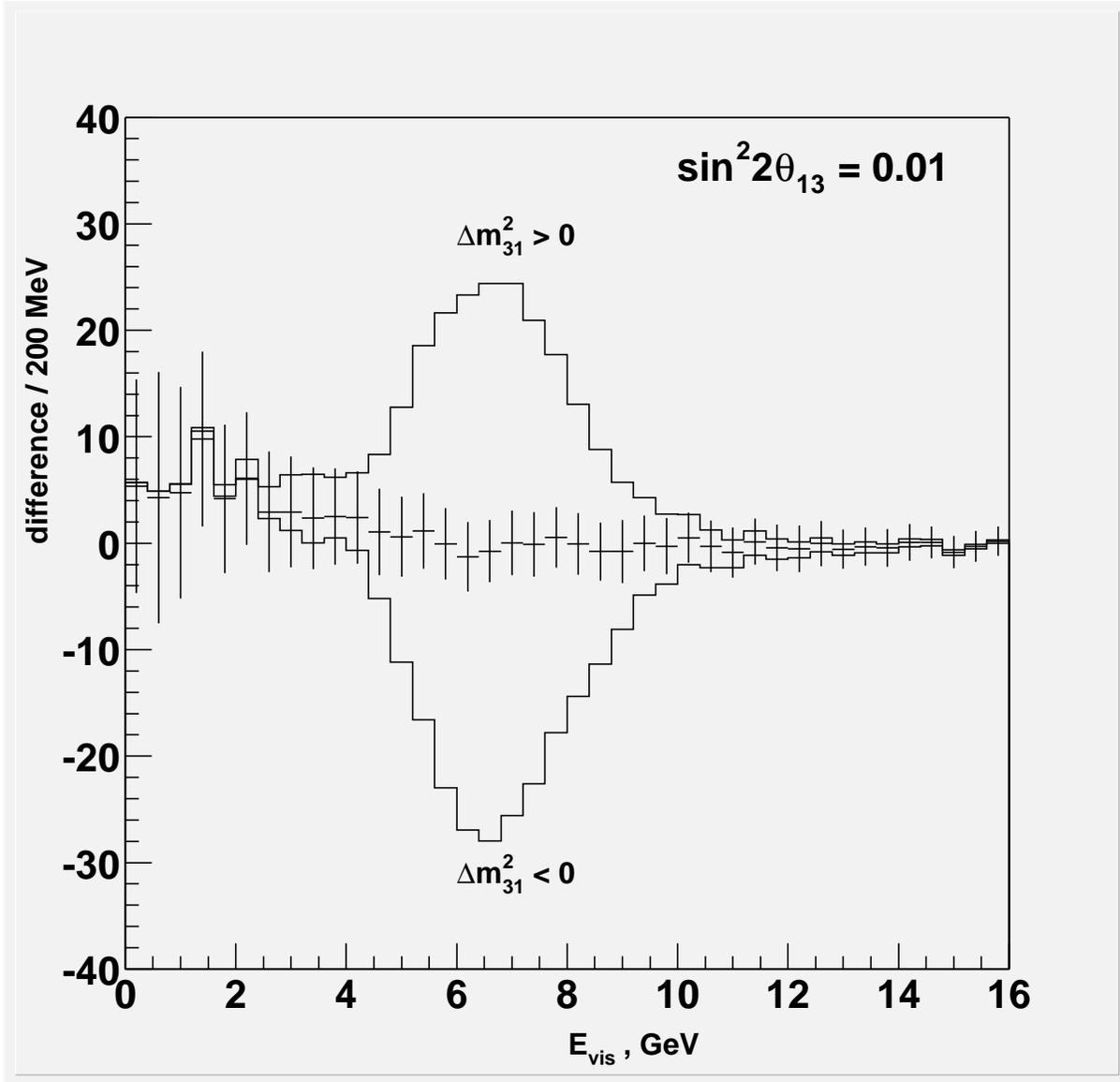}
\caption
{The difference between the \evis\ distributions for the $\nu$ and
$\bar{\nu}$ settins of the beam, assuming unequal $\nu$ and 
$\bar{\nu}$ exposures of 1.0 and 1.8 Mton$\times$yr, respectively.
The upper and lower histograms are for  $\Delta m^2_{31} > 0$ and
$\Delta m^2_{31} < 0$, respectively. The expectation for 
$\sin^2 2\theta_{13} = 0$ is illustrated by points with error 
bars that depict the statistical uncertainty.}
\label{difference}
\end{figure}

     The experimental strategy we adopt is to share the 
overall exposure between the $\nu$ and $\bar{\nu}$ running 
so as to equalize the expected backgrounds under the \omue\ and 
\omuebar\ signals, and then analyze the difference between the \evis\ 
distributions for the $\nu$ and $\bar{\nu}$ beams. The motivation is 
that many systematic uncertainties on the background should cancel 
out in the difference\footnote{This is 
particularly important here, as the large 
dip angle of the neutrino beam ($\sim 35^0$)
will rule out the construction of a "near" 
water Cherenkov detector.}.
The $\nu$ and $\bar{\nu}$ backgrounds
are approximately equalized by running 1.7--1.8 times longer in the 
$\bar{\nu}$ mode than in the $\nu$ mode (see Fig. \ref{together}).
The difference between the \evis\ distributions for the $\nu$ and 
$\bar{\nu}$ beams, assuming $\nu$ and $\bar{\nu}$ exposures of 1.0 and 
1.8 Mton$\times$yr, is illustrated in Fig. \ref{difference}. Depending 
on the sign of $\Delta m^2_{31}$, this distribution shows either a bump
or a dip at oscillation maximum with respect to the background that
corresponds to $\sin^2 2\theta_{13} = 0$.

     In order to estimate the significance of the oscillation signal in 
Fig. \ref{difference}, we vary the \evis\ interval so as to maximize
the "figure of merit" 
$F = (S_\nu - S_{\bar{\nu}}) / \sqrt{B_\nu + B_{\bar{\nu}}}$.
Here, $S_\nu$ and  $S_{\bar{\nu}}$  are the numbers of \omue\ and 
\omuebar\ events falling within the \evis\ interval, and $B_\nu$ and 
$B_{\bar{\nu}}$  are corresponding numbers of background events.
We obtain $F = +19.6$ for $\Delta m^2_{31} > 0$, and $F = -20.8$ for 
$\Delta m^2_{31} < 0$. Recalling that these figures refer to
$\sin^2 2\theta_{13} = 0.01$, we estimate that at 90\% CL the 
sensitivity to either $\sin^2 2\theta_{13}$ and the sign of 
$\Delta m^2_{31}$ will be maintained down to 
$\sin^2 2\theta_{13} \simeq 8\times10^{-4}$. 
Still lower values of $\sin^2 2\theta_{13}$ may perhaps be probed
with a neutrino factory in combination with a magnetized 
iron--scintillator detector \cite{factory, magic}. Note however that the
experimental scheme proposed in this paper is based on proven technology
and involves a multi-purpose facility \cite{detectors} rather than a
dedicated detector.

     To summarize, we have examined the physics potential of an
experiment with a neutrino superbeam that irradiates a Megaton-scale
water Cherenkov detector over the "magic" baseline $\sim$7300 km.
With realistic beam intensity and exposure, the experiment may 
probe $\sin^2 2\theta_{13}$ and the sign of $\Delta m^2_{31}$ down
to $\sin^2 2\theta_{13}$ values below $10^{-3}$. Thus obtained values 
of these parameters, that are not affected by degeneracies, can then 
be used as input for extracting $\delta_{CP}$ from the data collected 
with a shorter baseline as in the JHF--Kamioka experiment \cite{jhf2k}.

\clearpage

\end{document}